\documentstyle[12pt,aaspp4]{article}
\makeatletter

\@ifundefined{chapter}{\def\thebibliography#1{\section*{References\@mkboth
  {REFERENCES}{REFERENCES}}\list
  {\relax}{\setlength{\labelsep}{0em}
        \setlength{\itemindent}{-\bibhang}
        \setlength{\itemsep}{0pt}
        \setlength{\parsep}{0pt}
        \setlength{\leftmargin}{\bibhang}}
    \def\newblock{\hskip .11em plus .33em minus .07em}
    \sloppy\clubpenalty4000\widowpenalty4000
    \sfcode`\.=1000\relax}}%
{\def\thebibliography#1{\chapter*{Bibliography\@mkboth
  {BIBLIOGRAPHY}{BIBLIOGRAPHY}}\list
  {\relax}{\setlength{\labelsep}{0em}
        \setlength{\itemindent}{-\bibhang}
        \setlength{\itemsep}{0pt}
        \setlength{\parsep}{0pt}
        \setlength{\leftmargin}{\bibhang}}
    \def\newblock{\hskip .11em plus .33em minus .07em}
    \sloppy\clubpenalty4000\widowpenalty4000
    \sfcode`\.=1000\relax}}

\newlength{\bibhang}
\let\@internalcite\cite
\def\cite{\let\@citeleft(\let\@citeright)%
    \@ifstar{\citeyear}{\citefull}}
\def\acite{\let\@citeleft\relax\let\@citeright\relax%
    \@ifstar{\citeyear}{\acitefull}}
\def\citenp{\let\@citeleft\relax\let\@citeright\relax
    \@ifstar{\citeyear}{\citefull}}
\def\citefull{\def\astroncite##1##2{##1~##2}\@internalcite}
\def\citeyear{\def\astroncite##1##2{##2}\@internalcite}
\def\acitefull{\def\astroncite##1##2{##1~(##2)}\@internalcite}

\def\@citex[#1]#2{\if@filesw\immediate\write\@auxout{\string\citation{#2}}\fi
  \def\@citea{}\@cite{\@for\@citeb:=#2\do
    {\@citea\def\@citea{; }\@ifundefined
       {b@\@citeb}{{\bf ?}\@warning
       {Citation `\@citeb' on page \thepage \space undefined}}%
{\csname b@\@citeb\endcsname}}}{#1}}

\def\@cite#1#2{\@citeleft#1\if@tempswa , #2\fi\@citeright}
\def\@biblabel#1{}

\makeatother

\newcommand{\PSbox}[3]{\mbox{\rule{0in}{#3}\includegraphics{#1}\hspace{#2}}}
\newcommand{\FigNum}[1]{\unitlength 1pt \begin{picture}(55,10)(-400,35) 
                        \put(0,0){Figure #1}
                        \end{picture}}

\newcommand{\ppm}{\mbox{$\pm$}}

\def\etal{{et~al.}}

\newcommand{\nh}{\mbox{$N_{\rm H}$}}

\newcommand{\ud}[2]{\mbox{$^{+ #1}_{- #2}$}}
\newcommand{\ee}[1]{\mbox{$10^{#1}$}}
\newcommand{\tee}[1]{\mbox{$\times 10^{#1}$}}

\newcommand{\perval}[2]{{#1\mbox{$^{#2}$}}} 
\def\chisqr{\mbox{$\chi^2$}}
\def\dchisqr{\mbox{$\Delta \chi^2$}}
\def\chisqrnu{\mbox{$\chi^2_\nu$}}
\def\x1608{{4U~1608$-$522}}

\def\cenx4{{Cen~X$-$4}}

\def\saxj1808{{SAX J1808.4$-$3658}}

\newcommand{\chandra}{{\em Chandra\/}}

\newcommand{\xmm}{{\em XMM-Newton\/}}

\newcommand{\magnesium}{\ion{Mg}{11}}
\newcommand{\silicon}{\ion{Si}{14}}
\newcommand{\sulfur}{\ion{S}{16}}
\newcommand{\argon}{\ion{Ar}{18}}
\newcommand{\calcium}{\ion{Ca}{20}}
\newcommand{\rr}{{R02}}
\newcommand{\rrb}{{R02b}}
\newcommand{\grb}{{\small GRB}~011211}
\newcommand{\change}{ }

\begin{document}

\title{Statistical Re-examination of Reported Emission Lines in the X-ray Afterglow of GRB~011211}

\author{Robert E. Rutledge\altaffilmark{1} and Masao Sako\altaffilmark{1,2}}
\altaffiltext{1}{
Theoretical Astrophysics, 
California Institute of Technology, MS 130-33, Pasadena, CA 91125;
rutledge@tapir.caltech.edu, masao@tapir.caltech.edu}
\altaffiltext{2}{
Chandra Fellow}

\begin{abstract}
A 0.2-12 keV spectrum obtained with the \xmm\ EPIC/pn instrument of
\grb, taken in the first 5~ksec of a 27 ksec observation, was found by
Reeves \etal\ (2002; R02) to contain emission lines which were
interpreted to be from \magnesium, \silicon, \sulfur, \argon, and
\calcium, at a lower-redshift ($z_{obs}=1.88$) than the host galaxy
($z_{host}=2.14$).  We examine the spectrum independently, and find
that the claimed lines would not be discovered in a blind search.
Specifically, Monte Carlo simulations show that the significance of
reported features, individually, are such that they would be observed
in 10\% of featureless spectra with the same signal-to-noise.
Imposing a model in which the two brightest lines would be \silicon\
and \sulfur\ K$\alpha$ emission velocity shifted to between
$z$=1.88--2.40, such features would be found in between
$\sim$1.3-1.7\% of observed featureless spectra (that is, with
98.3-98.7\% confidence).  When we account for the number of trials
implicit in a search of five energy spectra (as were examined by \rr),
and permit a wider $z$-phase space search ($z=2.14$\ppm1.0), the
detection confidence of the two line complex decreases to 77-82\%.  We
find the detection significances to be insufficient to justify the
claim of detection and the model put forth to explain them.  $K\alpha$
line complexes are also found at $z=1.2$ and $z=2.75$ of significance
equal to or greater than that at $z=1.88$.  Thus, if one adopts the
$z=1.88$ complex as significant, one must also adopt the other two
complexes to be significant. The interpretation of these data in the
context of the model proposed by \rr\ is therefore degenerate, and
cannot be resolved by these data alone. Our conclusions are in
conflict with those of \rr, because our statistical significances
account for the multiple trials required -- but not accounted for by
\rr\ -- in a blind search for emission features across a range of
energies.  In addition, we describe a practical challenge to the
reliability of Monte Carlo \dchisqr\ tests, as employed by R02.
\end{abstract}


\section{Introduction}

It was recently reported that the X-ray afterglow of gamma-ray burst \grb, as
observed with \xmm\ EPIC/pn, contained spectral emission lines
\cite[hereafter, \rr]{reeves02} -- the first report of multiple X-ray emission
lines from a {\small GRB}.  These lines, at 0.45, 0.70, 0.89, 1.21, and 1.44
keV were interpreted to be from He-like \magnesium\ (rest energy 1.35 keV) and
H-like \silicon (2.0 keV), \sulfur\ (2.62 keV), \argon\ (3.32 keV) and
\calcium\ (4.10 keV), redshifted to $z$=1.88.  The difference between this and
the known redshift of the host galaxy $z_{host}=2.14$ was modeled as due to
supernova ejecta traveling at $v=25800$\ppm1200 km \perval{s}{-1}, which had
originated during a supernova 4 days prior to when the {\small GRB} jet
illuminated it, producing the afterglow (a more detailed analysis by
the same authors was completed after this paper was in its initial
form; \citenp[\rrb\ hereafter]{reeves02b}

The statistical significance of the individual lines was not reported
in \rr; it was stated that joint analysis of the lines taken together
produced an improvement in the \chisqr\ value which, by an F-test,
yielded a significance level of 99.7\%.  In addition, it was found
that Monte Carlo (MC) simulations were unable to produce the the same
improvement in \chisqr\ found between the best-fit power-law model and
the best-fit five emission-line model more than 0.02\% of the time.
Specifically, it was found that the best-fit \chisqr\ value for a
power-law model was improved by fitting to a model of a MEKAL plasma
with emission lines at rest energies corresponding to unresolved
\magnesium, \silicon, \sulfur, \argon, and \calcium\ redshifted to
$z=1.88$ in only 0.02\% of the simulated spectra (a 99.98\% confidence
detection).

The implications of the model discussed by \rr\ -- a delay between a
supernova and a {\small GRB} on a timescale of days, the formation of
a thin shell of supernova ejecta, an apparent under-abundance of Fe
relative to the detected nuclei -- provide severe constraints on
gamma-ray burst emission models.  In addition, as demonstrated by \rr,
the future detection of multiple emission lines can provide extremely
strong constraints on the production mechanisms, due to the inherent
required outflow velocity and emission timescales which can be derived
from them, not to mention the implied association with supernovae.
Similar spectra observed with greater S/N in the future would greatly
aid in unravelling the emission mechanisms and geometry of gamma-ray
bursts.  Therefore, it is of wide theoretical
(e.g. \citenp{lazzati02,kumar02}) and observational interest to
further interpret the observed X-ray spectrum of this \grb, in hopes
of determining what more could be learned from future, more precise
observations.

In Sec.~\ref{sec:anal}, we describe the observation, and perform a
basic spectral analysis using continuum models.  In Sec.~\ref{sec:mc},
we compare Monte Carlo (MC) realizations of acceptable continuum
models with the {\small GRB} spectrum, and find that the reported
features would be produced in $\sim$10\% of the continuum model
spectra, due only to Poisson noise.  In Sec.~\ref{sec:ka}, we adopt
the model that the two apparently most significant lines are K$\alpha$
lines of \silicon\ and \sulfur; we perform a blind search for features
of the same significance in MC realizations of continuum spectra, and
find that they would be reported from $\sim$1.2-2.6\% of such spectra,
again, due only to Poisson noise.  We describe in Sec.~\ref{sec:mcbad}
a practical challenge to the reliability of the MC \dchisqr\ analysis
produced by R02.  We conclude in Sec.~\ref{sec:con} that the lines are
not individually significant in the absence of an imposed model, and
are only marginally significant when the adopted model is imposed.

{\change\ 
These conclusions conflict with those of \rr. 
R02 derived the model (that is, the observed line energies, or
redshift) from the data, and then applied statistics for detection as
if the energies were known prior to examining the data (that is,
single-trial statistics). This is not appropriate when the model line
energies are derived directly from the X-ray data, and not from an
 a priori model -- one derived without examination of the X-ray data
(for example: line energies of multiple features with redshifts of the
host galaxy).  We adopt statistics appropriate to a blind-search for
these features, across a range of energies or redshifts (multi-trial
statistics).  This accounts for the diminished significance we find
for the features.
}
We further discuss the reasons for this conflict and conclude in
\S~\ref{sec:con}.

\section{Observation and Observed Spectrum }
\label{sec:anal}
\label{sec:spec}

We analyzed the identical source and background \xmm/EPIC-pn
\cite{struder01} spectrum as used by \rr\ (their Fig. 2), which was
kindly made available to us by the authors in electronic form
(J. Reeves, priv. comm.).  We used the same response matrix ({\tt
epn\_ff20\_sdY9\_thin.rsp}).  The spectrum used 5000 sec of realtime
observation beginning at 07:14:33 UT on 12 Dec 2001, with  a total
live time of 4440 sec.  The pn spectrum used counts comprised of
patterns 0--4 (singles and doubles), from a circular region 46\arcsec\
in radius centered on the source, excluding flagged events (for which
the keyword {\tt FLAG!=0}) and excluding a region near the edge of
the CCD chip. 

We performed a basic spectral analysis using XSPEC v11.1.0
\cite{xspec}.  We used data in the 0.2-12 keV energy range.  We
performed a non-standard spectral binning, implemented to maximize the
signal-to-noise associated with the reported emission features at the
reported energies.  We first binned data with energy bins centered at
the five best-fit line energies found by \rr, with bin-sizes
approximately equal to the FWHM EPIC/pn energy response at each energy
(respectively: 62 eV, 66 eV, 68 eV, 72 eV and 75 eV; see
Eq.~\ref{eq:fwhm}).  The remaining data were binned with 60 eV or
greater ($<$ 1 keV), and 70 eV or greater ($>$1 keV). Between 0.2 and
5 keV, each bin has $>$15 counts (although they were not binned on
this basis), for which $\chi^2$ fitting is valid.

We fit an absorbed photon power-law spectrum to the data, shown in
Fig.~\ref{fig:best}. The model spectrum was statistically acceptable (model
parameters are listed in Table~\ref{tab:obs}, along with the obtained \chisqr\
values).  We also fit the model with a thermal bremsstrahlung spectrum ({\tt
wabs*bremss}), and derived an acceptable best fit.  Finally, we found best-fit
model parameters for the values of the power-law photon slope (models 2 and 3)
and $kT_{\rm bremss}$ (models 5 and 6) at the 90\% confidence limits of the
best-fit, which will be used in MC simulations in \S~\ref{sec:mc}.

\section{Individual Emission Line Significances}

\label{sec:mc}
We first determine which of the reported lines are individually statistically
significant, when one is searching for emission lines at {\em a priori} known
energies.  We compared the observed spectrum with MC simulations of six
featureless spectra -- the three absorbed power-law and three absorbed thermal
bremsstrahlung which are models 1-6 in Table~\ref{tab:obs}.

We used a ``matched filter'', convolving the observed pulse-invariant (PI)
counts spectrum with a Gaussian energy response, with the energy resolution
response of the detector.  The matched filter approach maximizes the
signal-to-noise ratio as a function of energy of unresolved lines in the X-ray
PI spectrum.

Based on Figure 18 in \xmm\ v1.1 Users' Handbook
\cite{xmmusershandbook}, we modelled the photon energy redistribution
as a Gaussian response, with FWHM:

\begin{equation}
\label{eq:fwhm}
FWHM(E) = 57 + 13 (E/1 {\rm keV}) -0.29 (E/1{\rm  keV})^2 \; {\rm eV}
\end{equation}

\noindent This approximation was derived from the line in this figure.
The EPIC/pn energy resolution has been demonstrated to be stable over
9 months of in-flight calibration \cite{struder01}.  We expect
that this analysis (and that of \rr, since that work is based on the
same energy response matrices) is valid as long as the energy
resolution is within 20\% of this approximation (corresponding to 3 of
$\sim$15 PI channels at 0.75 keV).

We performed a convolution between the raw PI spectrum (that is,
number of counts vs. PI bin) and the gaussian energy response function, as a
function of energy:

\begin{equation}
\label{eq:c}
{\rm C(E_i)} = \sum_{j(E_i-3\sigma(E_i))}^{j(E_i+3\sigma(E_i))} I(j) \;  \frac{1}{\sqrt{2\pi}\sigma(E_i)}
\exp^{ -\frac{1}{2} \left(\frac{E_i-E_j}{\sigma(E_i)}\right)^2} \;
\delta E_j
\end{equation}

\noindent where $N$ is the number of PI bins, and we sum across PI
bins which are within \ppm3$\sigma(E_i)$ of $E_i$. $I(j)$ is the raw
PI spectrum, which contains both source and background counts, and
$j=1,2,..,N$ is the PI bin number.  The centroid (average) energies
and energy widths ($\Delta E_j$) of the PI bins were taken from the
{\tt EBOUNDS} extension of the response matrix, where $i$ is the PI
bin number and $\sigma(E)=FWHM(E)/2.35$.  We do not correct the PI
spectrum for the detector area; however the detector area does not
change dramatically across the FWHM of the lines.  If the area did
change dramatically across the FWHM of a line, and a statistical
excess were observed in the area-corrected PI spectrum but not in the
raw PI spectrum, then such an excess could well be be due to
calibration uncertainties.

The resulting $C(E_i)$ is shown in Fig.~\ref{fig:convolve}a.  By visual
inspection, there are indeed features in the spectrum near energies where the
reported lines occur.

To determine if these features are significant, we produced MC spectra of
models 1-6 (see \S~\ref{sec:spec}).  The MC realizations of the raw PI spectra
were performed as follows.  We simulated the spectral models 1-6 in XSPEC,
using the same response matrix as above, so that the resulting PI spectra
(without Poisson noise added) were convolved as the observed spectrum through
the telescope and detector response.  The simulated PI spectra N($E$) each had
a total of $>$9\tee{8} counts in PI bins between 0.2-3 keV.  We then produced
integrated spectra $I(E)=\int_{0.2\, \rm  keV}^E \, N(E) \, dE/\int_{0.2
\, \rm keV}^{3\,  \rm keV}
\, N(E) \,dE$, so that $I(0.2\,  {\rm keV})=0$ and $I(3\,  {\rm keV})=1$ (the integrated
normalized model is used for the MC simulation as described below).  These
constitute our six acceptable featureless spectral models; we will compare the
data with results from all six, as a firm conclusion that emission lines are
present should be independent of the underlying broad-band model assumed.

We implemented a background spectral model, to simulate the $\sim$10\% of the
counts due to background. Taking background from a different part of the
detector, we find that it can be parameterized by a broken photon power-law
({\tt bknpower}), with $\alpha_1=2.4$ at low energies, break energy 1.35 keV,
and $\alpha_2$=0.44 at high energies, between 0.20-7.3 keV (there is a strong
background line at 8 keV).  In fact, there are statistically significant
deviations from this pure continuum model between 0.55-0.6 keV; we ignore
these deviations in our background model.  In our MC simulation the effect of
ignoring what would appear to be a line in the observed spectrum is
conservative, in the sense that by ignoring its presence in the background
model, we could detect as ``significant'' a line in the 0.55-0.6 keV range
which is in fact produced by instrument background.

We simulated spectra between 0.2 and 3 keV, in which there were 560
counts in the observed spectrum, of which we estimate $\sim$66\ppm1.2
counts are due to background.  We drew, for each MC realization, a
number of background counts which is random poisson deviate (using
{\tt poidev}, \citenp{press}) with an average of 66 counts, with the
remaining (of 560) counts from the source. To produce a simulated
spectrum, we generate a random uniform deviate $r$ between 0 and 1,
and we place a count in the PI bin in which $I(E)=r$.

To produce our confidence limits to $C(E)$, we produced 1667 MC
realizations each of models 1-6 for a total of 10002 realizations.  We
set the 99\% and 99.9\% confidence limits at the 100th and 10th
greatest values, respectively, of $C(E)$ of all such realizations.
This insures that the conclusions are not dependent upon the assumed
featureless spectral model.

The results of this MC simulation are shown in
Fig.~\ref{fig:convolve}a.  Two of the reported features (near 0.7 keV
and 0.85 keV; \silicon\ and \sulfur) have single-energy-trial
probabilities of $>$99\% confidence in comparison with the featureless
spectral models (the claimed \sulfur\ line peaks just below the 99.9\%
confidence limit; we will treat it as having met 99.9\% confidence,
while the reader may regard this as an upper-limit).  The remaining
three lines are not significant in comparison with single-energy-trial
probability of 99\% confidence.

In Fig.~\ref{fig:convolve}, we also show $C(E)$ for single MC
realizations of the best-fit power-law spectrum (model 1), which also
contain apparent features.  The bumps in the single simulated spectra
appear because in any spectrum which contains Poisson noise, the
counts will not be distributed uniformly in energy, but will be
clustered in energy simply due to counting statistics.

\subsection{Multi-Energy-Trial (Blind Search) Probabilities}

Since it was necessary to perform a blind-search for emission line features in
the {\small GRB} spectrum -- as the redshifted line energies were not known
{\em a priori}, but were measured from the data -- it is necessary to estimate
the chance probability that the reported features are produced from a
featureless spectrum during a blind search for such features.

We produced 10000 MC realizations for each of models 1-6 as described in the
previous section.  We compared the $C(E)$ of these between 0.4 and 1.5 keV
against the single-energy-trial 99\% and 99.9\% confidence limits we found in
the previous section, for the models 1-6 individually.

In Table~\ref{tab:counts} we list the fraction of the 10000 MC spectra
in which $C(E)$ in at least one PI bin reaches a single-energy-trial
probability of 99\% or 99.9\% confidence. These fractions are 78-79\%
and 14-17\%, respectively; if finding a single-energy-trial 99\%
(\silicon) and 99.9\% (\sulfur) feature were statistically
independent, then the probability of observing both a 99\% and 99.9\%
single-energy-trial ``line'' in a single spectrum, such as we find in
the present spectrum of \grb, is $\approx$10\%.

Therefore, in a blind-search of the EPIC/pn spectrum for emission features, we
would expect to find features which have single-energy-trial significance
equal or greater to those observed in one of approximately ten observed
featureless spectra.

\section{Line Complex Significance As a Function of Redshift}
\label{sec:ka}

In this section, we examine if the reported lines, taken together, implicate
K$\alpha$ emission features from the particular redshift of $z$=1.88 as
reported by \rr.  We do so by summing the ${\rm C(E_i/(1+z))}$, using the
values of the rest energies of the reported lines:

\begin{eqnarray}
\chi(z) & = & \sum_i^{N_{\rm lines}} \sum_j^N I(j) \;  \frac{1}{\sqrt{2\pi}\sigma(E_j)}
\exp{ -\frac{1}{2} \left(\frac{E_j-(E_{\rm line,
i})/(1+z)}{\sigma(E_j)}\right)^2}\\
\end{eqnarray}

\noindent where $j$ denotes the PI bin number, $E_j$ is the centroid energy of
the $j$th PI bin, $i$ denotes the index $[1,5]$ of the five lines reported
detected; $E_i$ denotes the rest energies of the five lines identified by \rr,
which were 1.35 (\magnesium), 2.00 (\silicon), 2.62 (\sulfur), 3.32 (\argon),
and 4.10 keV (\calcium).  We examined the range of $0<z<3$, with a step-size of
$\Delta z=0.015$.  We use PI bins with energies 0.1-7 keV, to cover the
spectrum past the rest frame energy of \calcium.  We find 638 counts in this
energy range, of which we estimate 80 are from background.  We use only bins
which are within 3$\sigma(E_i/(1+z))$ of each $E_i/(1+z)$.  The result of this
convolution, if the reported lines are real, should be a maximum in $\chi(z)$
near the optimal redshift value, in excess of that found from MC 
realizations of data with featureless spectra.

The average value of $\chi$ will systematically increase with
increasing $z$ as the lines are shifted to lower energies, where the
intensity is higher in the power-law spectrum and the detector
effective area is larger and so there are a greater number of counts.
To examine if any particular maximum in $\chi(z)$ is significant, we
performed this convolution for 10000 MC realizations using the
simulated spectral models 1-6, taking the 100th and 10th highest
values, as described in the previous section, to produce the 99\% and
99.9\% confidence limits respectively. 

The results of the calculation using all 5 reported lines, as well as
the 99\% and 99.9\% MC confidence limits, are in
Fig.~\ref{fig:fivelines}.  The value of $\chi(z)$ is in excess of the
99\% MC confidence limit at $z=[1.86-1.98]$ and $z=[2.62-2.865]$, and
in excess of the 99.9\% MC confidence limit at $z=2.76$.

We also performed this convolution and Monte-Carlo simulation using
what appear to be the most significant two lines from Fig.2 of \rr\
(\silicon\ and \sulfur), the results of which are also shown in
Fig.~\ref{fig:fivelines}.  The value of $\chi(z)$ is in excess of the
99\% MC confidence limit at $z=[1.155-1.275]$ and $z=[1.80-2.01]$, and
in excess of the 99.9\% confidence limit at $z=[1.86-1.95]$.

\subsection{Multi-Redshift-Trial (Blind Search) Probabilities}

What fraction of featureless spectra, with the same number of source and
background counts as the observed spectrum, would produce values of $\chi(z)$
of comparable significance to the excess in $\chi(z=1.88)$ from the observed spectrum?  If one
examines $\chi(z)$ only at $z=1.88$, the answer is $<$1\%, which is the
single-$z$-trial probability.  However, the reported $z=1.88$ is different
from the known redshift of the host galaxy $z=2.14$; it is therefore unlikely
that $z=1.88$ was the only redshift which would be considered consistent with
an {\em a priori} model by \rr.  The pertinent statistical question to ask,
then, is what is the fraction of featureless X-ray spectra, examined for a
redshifted pair of \sulfur\ and \silicon\ lines, would produce a value of
$\chi(z)$ comparable to the single-$z$-trial significance observed, allowing
for a blind-search at values of $z$ between 1.88 and 2.40 (a range of equal
magnitude redshift and blueshift from the host galaxy)?

To address this, we simulated 10000 MC spectra of each of spectral
models 1-6, and found $\chi(z)$ in the same way as for the observed
spectrum in the previous section.  We used only the 2-line model, as
this gave the apparently most significant result near $z=1.88$.  We
compared $\chi(z)$ with the 99\% and 99.9\% MC limits, found in the
previous section, and noted when these were exceeded in at least one
$z$ bin for the 99\% confidence limit, and in at least seven
consecutive $z$ bins for the 99.9\% confidence limit between $z=1.88$
and $z=2.40$.  We require seven consecutive $z$ bins as this is the
number of $z$ bins in $\chi(z)$ we find in excess of the single-trial
99.9\% confidence limit near $z=1.88$.  (We require only 1 bin for the
99\% confidence limit to satisfy a minimal ``detection'' requirement;
whereas we require seven bins for the 99.9\% confidence limit, since
this was what was actually observed near $z=1.88$, and we wish to
evaluate the likelihood of producing the observed $\chi(z)$ excess).

The fraction of MC featureless spectra which contained at least 1 $z$
bin between $z=1.88$ and $z=2.40$ in excess of the single-$z$-trial MC
probability of 99\% are given in Table~\ref{tab:ztab}.  Because the
observed spectrum gave $\chi(z)>99.9$\% in seven consecutive $z$ bins,
we also used this as our criterion to count ``hits'' in the $>$99.9\%
confidence comparison, also shown in Table~\ref{tab:ztab}. Of 10000 MC
spectra, between 20-22\% produced ``hits'' for the single-$z$-trial
99\% confidence limit, and 1.5-1.9\% produced ``hits'' for the the
single-$z$-trial 99.9\% confidence limit.

We note that when we search the range $z=2.14\pm1.0$ (instead of
\ppm0.26) the percentage of featureless spectra which have seven
consecutive $z$ bins with $\chi(z)$ greater than the 99.9\% limit is
3.8-5.0\%. However, it is unclear if \rr\ would have attached equal
significance to a detection at $z=1.14$ as one at $z=1.88$, as no
limits on excess line emission as a function of assumed redshift are
given, and the redshift phase-space examined by \rr\ was not given.
We therefore rely on our search of the smaller phase-space; while this
may underestimate the number of "effective trials" used by \rr, it
nonetheless serves as the probability of producing the claimed excess
line emission due a statistical fluctuation within the $\delta z=0.26$
observed.  If the redshift space examined by \rr\ were 1.14-3.14
($\delta z=1.0$), then the probability of finding an excess equal or
greater than that observed would be 3.8-5.0\%.  If the full redshift
space of 0-5 was in fact examined by \rr\, then the probability of a
false detection is $>$3.8-5.0\%.

\section{A Practical Challenge with Monte Carlo \dchisqr\ Tests for
Multi-Parameter Models}
\label{sec:mcbad}

A MC \dchisqr\ test as employed by R02 is not fundamentally flawed as
is the analytic \dchisqr\ test (that is, the F-test) for the
application of spectral emission line discovery.  In the F-test, the
reference \dchisqr\ distribution was derived under the assumption that
the null hypothesis lies on the border of the acceptable parameter
space \cite{protassov02}, which is not true in a search for emission
lines; however, this assumption is not made in the MC \dchisqr\ test.
Thus, the simulated \dchisqr\ distribution can, in principle, provide
a reliable reference distribution with which the value of \dchisqr\
from application to real data can be compared to determine the false
positive rate.

However, as we show below, the MC \dchisqr\ test as employed by \rr\
(and described more fully by \rrb) suffers from a practical problem
which makes it an inferior approach to the one we have applied.
Specifically, to apply the \dchisqr\ statistic using the MC approach,
one must assuredly find the {\em global} minimum \chisqr\ for the
applied model for every single MC realization; the description of the
analysis performed by R02 (and \rrb) does not assure that this has
occurred.

In the case of \chisqr\ minimization through local mapping of the
\chisqr\ surface, as in the modified Levenberg-Marquart method
employed in XSPEC (\citenp{xspec}; modified from the CURFIT algorithm
as described by \citenp{bevington}; see also \citenp{press}), one
finds the vector in multi-parameter space along this surface which
provides the most negative derivative, follows along this vector a
short way, and iterates, until one reaches a point where there are no
negative derivatives in any direction along the \chisqr\ surface (that
is, when one has reached a minimum point.  This approach suffers from
the well known problem of local minima, where the true global minimum
can lie at a completely different set of parameter values (see, for
example \citenp{press}, p. 394).  On simple \chisqr\ surfaces, where
the second partial derivatives of the \chisqr\ surface are everywhere
small -- certainly in the case of the 2- parameters power-law spectrum
-- it is rare that local minima different from the global minimum are
found.  However, on complex \chisqr\ surfaces (those which contain
large second partial derivatives of \chisqr) -- as will be the case
when fitting a six parameter model of three emission lines of
specified rest energy with variable fluxes and redshift plus a
power-law (slope and normalization) -- local minima are common;
subsequently, this approach is unsuited to the unassisted discovery
(by computer alone, without human intervention) of the global \chisqr\
minimum.  It is, for example, common occurrence when using a
multi-component (of more than, say, three) parameters in XSPEC that
some final human assistance is required to find the global minimum,
since almost always it is a local minimum which is found unassistedly
by the computer; the quantitative difference in \chisqr\ between the
computer-discovered local minimum and the true global minimum will
depend on the complexity of the \chisqr\ surface.  In general,
\chisqr\ surfaces become more complex with the addition of more model
parameters, and the discrepancy will be greater when there are greater
covariances between model parameters (such as can be expected between
flux for line 1 vs. line 2, or vs. the continuum, or for each of the
lines and the continuum, or for the relative flux for the lines and
the spectral slope; and so on).  The only certain means to overcome
this known deficiency is to evaluate \chisqr\ on a parameter grid
with resolution in each parameter dimension much smaller than ranges
where the \chisqr\ value changes by 1.

Thus, while the global minimum will likely be found from unassisted
discovery for the power-law spectral model, it is more likely that
only a local minimum will be found for the six-parameter model when
the search for this minimum is not human-assisted.  This will
underestimate the value of \dchisqr\ for that realization; over the
entire ensemble of MC realizations, there are then fewer false
positives, and the significance of the \dchisqr\ in support of the
presence of lines will be overstated.  In addition to the problem of
local minima, the local minimum found will be dependent upon initial
parameter values (that is, the algorithm is path-dependent); it 
therefore does not lend itself to duplication by different
groups. Also, the spectral fitting is non-analytic, such that the
magnitude of a possible discrepancy cannot be evaluated {\em a
priori}.

XSPEC -- which R02 states was used for the MC simulation -- does
perform the \chisqr\ minimization approach.  We suggest that it is
unlikely that human-assisted spectral fitting -- as is common practice
when attempting to find the global \chisqr\ minimum for a single
spectrum in XSPEC -- was performed for all 10,000 MC spectra by R02,
as we found ourselves was necessary for the single spectral fit to the
real data, as this would be an impractically long task.

We are unable to attempt to duplicate the result of R02, because
performing assisted spectral fitting on 10,000 MC spectra is
impractical, and in any case, we know of no deficiency in our present
approach.  Our approach, in contrast, is analytic and not
path-dependent and, therefore, more robust.

\section{Discussion and Conclusions}
\label{sec:con}

We have attempted to confirm the observational statistical
significance of emission lines in the X-ray afterglow of \grb.  In a
blind-search for individual emission lines between 0.4 and 1.5 keV,
features of significance equal to those observed will be found in one
in ten featureless spectra.  Thus, the reported features can be said
to be detected with 90\% confidence in a model-independent way.

Also, a blind-search for the two-line complex (\silicon\ and \sulfur)
at any redshift between the reported value ($z=1.88$) and a blueshift
of equal magnitude from the host galaxy ($z=2.40$) would find such
features with equal significance to that observed in 1 of 60
featureless spectra (1.3-1.7\% of the time, depending on the intrinsic
spectrum).  Thus, the features as reported can be said to be detected
with 98.3-98.7\% confidence, in a model-dependent interpretation,
where we search for two features due to \silicon\ and \sulfur\
K$\alpha$ redshifted to some value of $z$ in the range
$z=$2.14\ppm0.26.

The difference between the present statistics and those of \rr\ are
due to the different statistical arguments used to establish the
existence of the emission lines.  While \rr\ relies on single-trial
statistics, we find the model used by \rr\ (K$\alpha$ lines, at a
redshift different from that of the host galaxy) was derived directly
from the data, which therefore requires a statistical analysis
appropriate to a blind search.  By expanding the searched phase-space,
and taking into account the multiple trials of a blind search, the
confidence in the detection drops from the 99.98\% of \rr\ to, the
98.7\% (best case) we find here.  Moreover, \rr\ did not estimate the
individual significances of the lines as we do here; thus we find that
such ``lines'' would appear in between 15-78\% of observed featureless
spectra for a single-trial significance comparable to that of the
reported \silicon\ or \sulfur\ lines. 

{ The analysis of these data has otherwise recently been
called into question.  \acite{lanl02} have shown that there is a
background line associated with the EPIC/pn detector edge during the
observation, which would have been included in the {\small GRB}
spectrum from the first 5 ksec, when the source was near the detector
edge, but not afterwards, after the source had been moved away from
the detector edge.  In our own analysis, we cannot confirm this result
unless we adopt non-standard event selection criteria, which differ
from the ones used by \rr.  \rr\ removed events near the CCD detector
edge ({\tt FLAG==0}) and selected only single and double events ({\tt
PATTERN<=4}) (J. Reeves, priv. comm.).  These selections result in a
smooth, featureless background spectrum with with no bright line-like
feature near $\sim 0.7 ~\rm{keV}$ (as seen in Fig.~4f of
\citenp{lanl02}) as well as a reduction of the count rate by a factor
of $\ga 2$ in the range $E = 0.2 - 3 ~\rm{keV}$ (see
Fig.~\ref{fig:masao}).  Therefore, we are not able  to confirm the
applicability of \cite{lanl02} to the analysis of \rr.  
}

An alternative approach to the one we have taken is employed using
XSPEC, in which one fits a featureless spectrum to the data, and then
a spectrum which includes emission lines, to determine if the change
in \chisqrnu\ is significant, as according to an F-test; this is the
approach taken by \rr.  However, this approach for the detection of
emission or absorption lines is formally incorrect, and gives false
statistical results \cite{protassov02} particularly so when the true
continuum is not well constrained, as in the present case.  We
therefore prefer our approach of applying a matched energy response
filter for line detection at arbitrary energies, and to compare this
with application of the matched filter to MC realizations of featureless
spectra.  It is a trivial statistical exercise to demonstrate that
matched filtering maximizes the signal-to-noise ratio (and thus
detectability) for detection of infinitely narrow emission lines.

In estimating the model-dependent confidence limit for the detection
of the line complex (98.7\%), we accounted only for searching the
redshift phase space between $z=1.88$ and $z=2.40$, symmetric about
the host galaxy redshift -- an extremely minimal requirement.  We did
not account for the full redshift phase space searched by \rr , as
such was not given in that reference; if the redshift phase space
searched by \rr\ covered $z=1.14-3.14$ ($z=2.14\ppm1.0$), then the
detection significance of the two strongest lines (\silicon\ and
\sulfur) together decreases from 98.3-98.7\% to 95-96.2\% confidence.
Finally, we did not include in this confidence limit the number of
trials implicit in searching five X-ray spectra for emission lines,
which was performed by \rr\ for different time periods (0-5 ksec, 5-10
ksec, 10-15 ksec, 15-20 ksec, and 20-27 ksec).  If we presume the same
search was made on all five spectra, as seems a reasonable {\em a
priori} search to perform, then the detection confidence for the
\silicon\ and \sulfur\ lines together decreases to
$0.95^5$--$0.962^5$=77-82\%.  We regard 98.7\% to be a conservative
(in the sense of permitting a higher significance) upper-limit to the
confidence of detecting the \silicon\ and \sulfur\ lines together,
while a more accurate accounting of the number of trials and
phase-space searched by \rr\ produces a 77-82\% confidence limit.

We consider neither a 90\% confidence detection in a model-independent
interpretation, nor a 98.3-98.7\% confidence detection in a
model-dependent interpretation, to be sufficient to justify the
detection claims and subsequent interpretation put forth by \rr.  The
77-82\% confidence limit, which accounts for the wide $z$-phase space
and number of spectra examined by \rr, is well below any comfortable
detection confidence. If the $z$ phase space actually searched by \rr\
is larger, the number of implicit trials is greater, and our estimate
of the confidence level for the detected line complex would decrease.

Moreover, if one concludes that the marginal detection of the 2 lines
(Si \& S) near $z=1.88$ is significant, then one must also conclude
that the detection of all 5 lines near $z=2.75$ is equally
significant.  In addition, if one concludes that the marginal
detection of the 5 lines near $z=1.88$ is  significant,
then one must also conclude that the detection of 2 lines (Si \& S)
near $z=1.2$ is equally significant. 

Therefore, one cannot conclude simply that a complex of K$\alpha$ line
emission is detected near $z=1.88$; these data permit alternate
interpretations of such complexes near $z=1.2$ and $z=2.75$.  As the
statistical excesses are due to the same ``features'' in the observed
spectrum, the interpretation of the statistical excess in the context
of the model presented by \rr\ is degenerate and cannot be resolved
with these data alone.

Prospects for confirmation of line features in GRBs are very good,
considering that the X-ray spectral integration for \grb\ was begun 11
hours after the {\small GRB} was initially detected, and required 1.4
hrs of integration to obtain.  Decreasing the reaction time would
permit a longer integration, while the afterglow is brighter in the
X-rays, and the marginal results found here may well be improved upon.

\acknowledgements

We are grateful to J. Reeves, who generously made his observed
spectrum of the first 5 ksec of the \xmm\ observation of \grb\
available to us, that we might independently analyze it.  We
gratefully acknowledge useful conversations with A. MacFadyen,
R. Blandford, and D. Fox.  The authors are grateful to F. Harrison,
F. Paerels and an anonymous referee for useful comments on the
manuscript. {\small MS} was supported by {\small NASA} through {\it
Chandra} Postdoctoral Fellowship Award Number {\small PF}1-20016
issued by the \chandra\ X-ray Observatory Center, which is operated by
the Smithsonian Astrophysical Observatory for and behalf of {\small
NASA} under contract {\small NAS}8-39073.

\clearpage

\begin{figure}[htb]
\caption{ \label{fig:best} ({\bf Top Panel}): X-ray spectrum from
\xmm\ EPIC/pn of \grb, with energy binning optimized to find
deviations from the best-fit power-law spectrum (solid line) due to
emission lines at the reported energies, with a best-fit absorbed
power-law model.  The data show no significant excess counts at the
reported line energies. ({\bf Bottom Panel}):
$\chi=$(model-data)/$\sigma$, residuals between the best-fit continuum
spectra and the data.  The 5-10 keV energy bin, while included in our
spectral fits, is not included in this figure, to better show the 0.2-5
keV energy spectrum. }
\end{figure}

\begin{figure}[htb]
\caption{ \label{fig:convolve} ({\bf Panel a}): Solid line is $C(E)$
(Eq.~\ref{eq:c}) from the observed raw PI spectrum -- the convolution
between the raw spectrum and the EPIC/pn energy response.  The broken
lines are the max($C(E)$) for spectral models 1-6, showing the 99\%
and 99.9\% confidence single-trial upper-limits.  ({\bf panels b-f}):
The solid  line is the same observed convolved spectrum as in Panel a.
Dotted  lines are five (in the five separate panels) randomly selected
Monte Carlo spectra using Model 1.  Features of similar magnitude to
those found in the observed spectra are apparent in each; these are
due to the Poisson noise distribution (in energy) in a spectrum with a
finite number of detected counts. }
\end{figure}

\begin{figure}[htb]
\caption{ \label{fig:fivelines} ({\bf Top panel}) Figure of merit
$\chi(z)$ using all five reported line energies (solid line), for
$0<z<3$.  We also show the extremum Monte-Carlo values for 99\% (long
dashed) and 99.9\% confidence (short dashed), using  all six model
spectra. The solid vertical line marks the redshift of the reported
detection (z=1.88); at this redshift, the $\chi(z=1.88)$.   ({\bf Bottom
Panel}) Same, except for only \silicon\ and \sulfur\ lines. Again,
the $\chi(z=1.88)$ is below the 99\% confidence limit.  }
\end{figure}

\begin{figure}[htb]
\caption{ \label{fig:masao}  The background spectrum near the region
near the CCD chip edge using two different event selection criteria.
(a) no explicit selection of {\tt PATTERN} and {\tt FLAG} as adopted
by Borozdin \& Trudolyubov (2002) (circles) binned at a minimum of 5
counts per bin and (b) using only {\tt PATTERN<=4} and {\tt FLAG=0}
events as adopted by \rr\ (stars) with binnsizes identical to those
of (a).}
\end{figure}

\clearpage
\pagestyle{empty}
\begin{figure}[htb]
\PSbox{fig1.ps hoffset=-80 voffset=-80}{14.7cm}{21.5cm}
\FigNum{\ref{fig:best}}
\end{figure}

\clearpage
\pagestyle{empty}
\begin{figure}[htb]
\PSbox{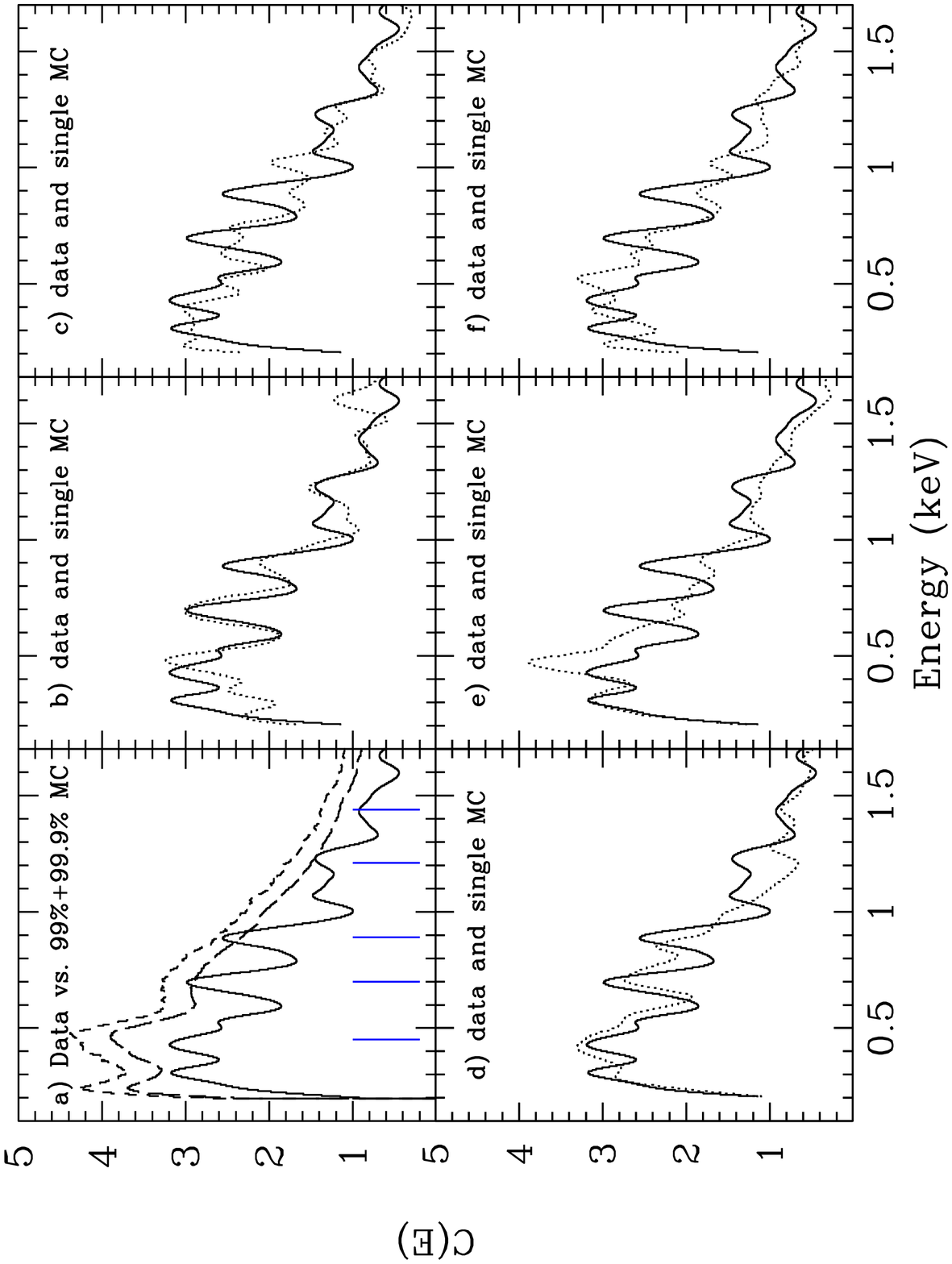 hoffset=-80 voffset=-80}{14.7cm}{21.5cm}
\FigNum{\ref{fig:convolve}}
\end{figure}

\clearpage
\pagestyle{empty}
\begin{figure}[htb]
\PSbox{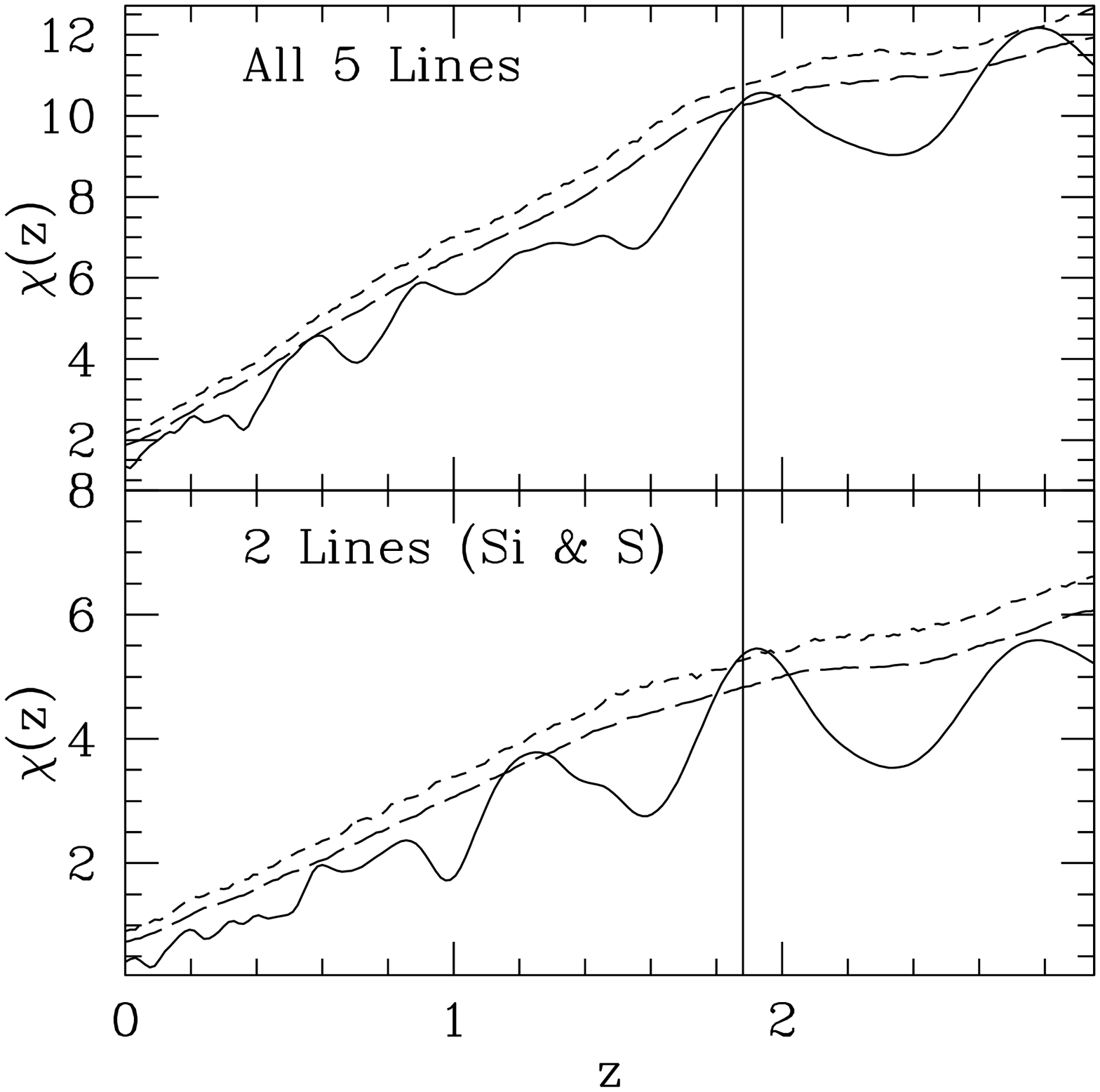 hoffset=-80 voffset=-80}{14.7cm}{21.5cm}
\FigNum{\ref{fig:fivelines}}
\end{figure}

\clearpage
\pagestyle{empty}
\begin{figure}[htb]
\PSbox{fig4.ps  hoffset=-80 voffset=-80}{14.7cm}{21.5cm}
\FigNum{\ref{fig:masao}}
\end{figure}

\clearpage

\begin{deluxetable}{lllll}
\tablecaption{\label{tab:obs} Best-fit X-ray Spectral Parameters}
\tablewidth{17cm}
\tablehead{
\colhead{} & 
\colhead{\nh} &
\colhead{} &
\colhead{N} &
\colhead{\chisqrnu/dof}\\
\colhead{Model} & 
\colhead{(\ee{22} \perval{cm}{-2})} &
\colhead{$\alpha$/$kT_{\rm bremss}$} &
\colhead{phot/keV/cm$^2$/s at 1 keV} &
\colhead{ (prob$^a$)}	
}
\startdata
(1) Best Fit 	& 0.10\ud{0.04}{0.02}	& 2.6\ppm0.2	& (7.0\ppm0.9)\tee{-5} & 1.24/17  (0.22) 	\\
2 		& 0.065\ppm0.01		& (2.3)		& (5.8\ppm0.5)\tee{-5} & 1.35/18  (0.14)  \\
3		& 0.16\ppm0.02		& (3.0)		&(8.5\ppm0.7)\tee{-5}  & 1.35/18  (0.15) \\ \hline
\multicolumn{5}{c}{Thermal Bremmstrahlung} \\
(4) Best Fit 	& 0.03\ppm0.01		& 1.5\ud{0.4}{0.2}& (9.0\ppm1.2)\tee{-5}    	& 1.38/17  (0.13)\\
5 		& 0.05\ppm0.01		& (1.1)		& (12\ppm1)\tee{-5}		& 1.50/18 (0.08) \\
6 		& 0.014\ppm0.01		& (2.1)		& (7.0\ud{0.7}{0.4})\tee{-5}	& 1.44/18 (0.10) \\
\enddata
\end{deluxetable}

\begin{deluxetable}{lll}
\tablecaption{\label{tab:counts} Fraction of Featureless MC Spectra which produce
single-energy-trial ``lines'' at 99\% and 99.9\% confidence between
0.4 and 1.5 keV}
\tablewidth{10cm}
\tablehead{
\colhead{Model} & 
\colhead{$>$99\% (1 $z$ bin)} &
\colhead{$>$99.9\% (1 $z$ bin)} 
}
\startdata
1 	 	& 0.78	& 0.16	\\
2 		& 0.78  &  0.15 \\
3		& 0.78  & 0.14 \\
4		& 0.78 & 0.15 \\ 
5 		& 0.79 & 0.16 \\
6 		& 0.78 & 0.17 \\
\enddata
\end{deluxetable}

\begin{deluxetable}{lll}
\tablecaption{\label{tab:ztab} Fraction of Featureless MC Spectra which produce
single-$z$-trial $\chi(z)$ for 2-lines at 99\% and 99.9\% confidence between
$z=1.88$ and $z=2.40$ }
\tablewidth{10cm}
\tablehead{
\colhead{Model} & 
\colhead{$>$99\% (1 $z$ bin)} &
\colhead{$>$99.9\% (7 $z$ bins)} 
}
\startdata
1 	 	& 0.20	& 0.015	\\
2 		& 0.20  &  0.014 \\
3		& 0.20  & 0.014 \\
4		& 0.21 & 0.015 \\ 
5 		& 0.20 & 0.013 \\
6 		& 0.22 & 0.017 \\
\enddata
\end{deluxetable}


\begin{thebibliography}{}

\bibitem[\protect\astroncite{Arnaud}{1996}]{xspec}
Arnaud, K.~A., 1996,
\newblock in G. Jacoby \& J. Barnes (eds.), {\em Astronomical Data Analysis
  Software and Systems V.}, Vol. 101, p.~17, ASP Conf. Series

\bibitem[\protect\astroncite{Bevington}{1969}]{bevington}
Bevington, P.~R., 1969,
\newblock {\em Data Reduction and Error Analysis for the Physical Sciences},
\newblock McGraw-Hill

\bibitem[\protect\astroncite{{Borozdin} \& {Trudolyubov}}{2002}]{lanl02}
{Borozdin}, K.~N. \& {Trudolyubov}, S.~P., 2002,
\newblock {\em \apjl},
\newblock submitted, astro-ph/0205208

\bibitem[\protect\astroncite{Dahlem}{1999}]{xmmusershandbook}
Dahlem, M., 1999,
\newblock {\it XMM Users' Handbook, Issue 1.1}, distributed by the XMM-Newton
  Science Operations Center, Vilspa

\bibitem[\protect\astroncite{{Kumar} \& {Narayan}}{2002}]{kumar02}
{Kumar}, P. \& {Narayan}, R., 2002,
\newblock {\em \apj},
\newblock submitted, astro-ph/0205488

\bibitem[\protect\astroncite{{Lazzati} {\rm et~al.\/}}{2002}]{lazzati02}
{Lazzati}, D., {Ramirez-Ruiz}, E., \& {Rees}, M.~J., 2002,
\newblock {\em \apjl} { 572}, L57

\bibitem[\protect\astroncite{Press {\rm et~al.\/}}{1995}]{press}
Press, W., Flannery, B., Teukolsky, S., \& Vetterling, W., 1995,
\newblock {\em Numerical Recipies in C},
\newblock Cambridge University Press

\bibitem[\protect\astroncite{{Protassov} {\rm et~al.\/}}{2002}]{protassov02}
{Protassov}, R., {van Dyk}, D.~A., {Connors}, A., {Kashyap}, V.~L., \&
  {Siemiginowska}, A., 2002,
\newblock {\em \apj} { 571}, 545

\bibitem[\protect\astroncite{{Reeves} {\rm et~al.\/}}{2002a}]{reeves02}
{Reeves}, J.~N., {Watson}, D., {Osborne}, J.~P., {Pounds}, K.~A., {O'Brien},
  P.~T., {Short}, A.~D.~T., {Turner}, M.~J.~L., {Watson}, M.~G., {Mason},
  K.~O., {Ehle}, M., \& {Schartel}, N., 2002a,
\newblock {\em \nat} { 416}, 512

\bibitem[\protect\astroncite{{Reeves} {\rm et~al.\/}}{2002b}]{reeves02b}
{Reeves}, J.~N., {Watson}, D., {Osborne}, J.~P., {Pounds}, K.~A., {O'Brien},
  P.~T., {Short}, A.~D.~T., {Turner}, M.~J.~L., {Watson}, M.~G., {Mason},
  K.~O., {Ehle}, M., \& {Schartel}, N., 2002b,
\newblock {\em \aa},
\newblock submitted, astro-ph/0206480

\bibitem[\protect\astroncite{{Str{\" u}der} {\rm et~al.\/}}{2001}]{struder01}
{Str{\" u}der}, L., {Briel}, U., {Dennerl}, K., {Hartmann}, R., {Kendziorra},
  E., {Meidinger}, N., {Pfeffermann}, E., {Reppin}, C., {Aschenbach}, B.,
  {Bornemann}, W., {Br{\" a}uninger}, H., {Burkert}, W., {Elender}, M.,
  {Freyberg}, M., {Haberl}, F., {Hartner}, G., {Heuschmann}, F., {Hippmann},
  H., {Kastelic}, E., {Kemmer}, S., {Kettenring}, G., {Kink}, W., {Krause}, N.,
  {M{\" u}ller}, S., {Oppitz}, A., {Pietsch}, W., {Popp}, M., {Predehl}, P.,
  {Read}, A., {Stephan}, K.~H., {St{\" o}tter}, D., {Tr{\" u}mper}, J., {Holl},
  P., {Kemmer}, J., {Soltau}, H., {St{\" o}tter}, R., {Weber}, U., {Weichert},
  U., {von Zanthier}, C., {Carathanassis}, D., {Lutz}, G., {Richter}, R.~H.,
  {Solc}, P., {B{\" o}ttcher}, H., {Kuster}, M., {Staubert}, R., {Abbey}, A.,
  {Holland}, A., {Turner}, M., {Balasini}, M., {Bignami}, G.~F., {La
  Palombara}, N., {Villa}, G., {Buttler}, W., {Gianini}, F., {Lain{\' e}}, R.,
  {Lumb}, D., \& {Dhez}, P., 2001,
\newblock {\em \aap} { 365}, L18

\end{thebibliography}
\end{document}